\documentclass[a4paper,12pt,epsfig]{article}
\usepackage{epsfig}
\usepackage{amssymb}
\usepackage{amsfonts}
\usepackage{amsmath}
\usepackage{euscript}
\usepackage{verbatim}
\usepackage{latexsym}
\usepackage{graphicx}

\newskip\humongous \humongous=0pt plus 1000pt minus 1000pt

\newif\ifdtup

\catcode`\@=11

\@addtoreset{equation}{section}

\def\@normalsize{\@setsize\normalsize{15pt}\xiipt\@xiipt
\abovedisplayskip 14pt plus3pt minus3pt%
\belowdisplayskip \abovedisplayskip
\abovedisplayshortskip \z@ plus3pt%
\belowdisplayshortskip 7pt plus3.5pt minus0pt}

\def\small{\@setsize\small{13.6pt}\xipt\@xipt
\abovedisplayskip 13pt plus3pt minus3pt%
\belowdisplayskip \abovedisplayskip
\abovedisplayshortskip \z@ plus3pt%
\belowdisplayshortskip 7pt plus3.5pt minus0pt
\def\@listi{\parsep 4.5pt plus 2pt minus 1pt
     \itemsep \parsep
     \topsep 9pt plus 3pt minus 3pt}}

\relax

\catcode`@=12

\catcode`\@=11

\def\section{\@startsection{section}{1}{\z@}{3.5ex plus 1ex minus
   .2ex}{2.3ex plus .2ex}{\large\bf}}

\def\SymBoxes#1#2#3#4{\newdimen\un@t \un@t#3%
\raisebox{#1}{\rule{#2\un@t}{#4}\hskip-#2\un@t% lower horizontal
\@tempdimb\un@t \advance\@tempdimb by-#4\@tempcntb#2\relax%
\@whilenum{\@tempcntb>0}\do{%                         % #2 vertical lines
\rule{#4}{\un@t}\hskip\@tempdimb \advance\@tempcntb by\m@ne}%
\hskip-#2\un@t \rule[\un@t]{#2\un@t}{#4}%
\rule[\un@t]{#4}{#4}\hskip-#4%             % upper horizontal line
\rule{#4}{\un@t}}\hskip-#4}                % rightest vertical line
\begin{document}
%\begin{letter}{~}

%%%%%%Define some new commands and  macros
\newcommand{\beq}{\begin{equation}}
\newcommand{\eeq}{\end{equation}}
\newcommand{\bea}{\begin{eqnarray}}
\newcommand{\eea}{\end{eqnarray}}
\newcommand{\beas}{\begin{eqnarray*}}
\newcommand{\eeas}{\end{eqnarray*}}
\newcommand{\defi}{\stackrel{\rm def}{=}}
\newcommand{\non}{\nonumber}
\newcommand{\bquo}{\begin{quote}}
\newcommand{\enqu}{\end{quote}}
%%%%%%%%%%%%%%%%
\renewcommand{\(}{\begin{equation}}
\renewcommand{\)}{\end{equation}}
%%%%%%%%%%%%%%%%%%%%%%%%%%%%%%%%%% definitions
\def \eqn#1#2{\begin{equation}#2\label{#1}\end{equation}}
\def\IZ{{\mathbb Z}}
\def\IR{{\mathbb R}}
\def\IC{{\mathbb C}}
\def\IQ{{\mathbb Q}}
\def\de{\partial}
\def\Tr{ \hbox{\rm Tr}}
\def\H{ \hbox{\rm H}}
\def\HE{ \hbox{$\rm H^{even}$}}
\def\HO{ \hbox{$\rm H^{odd}$}}
\def\K{ \hbox{\rm K}}
\def\Im{ \hbox{\rm Im}}
\def\Ker{ \hbox{\rm Ker}}
\def\const{\hbox {\rm const.}}
\def\o{\over}
\def\im{\hbox{\rm Im}}
\def\re{\hbox{\rm Re}}
\def\bra{\langle}\def\ket{\rangle}
\def\Arg{\hbox {\rm Arg}}
\def\Re{\hbox {\rm Re}}
\def\Im{\hbox {\rm Im}}
\def\exo{\hbox {\rm exp}}
\def\diag{\hbox{\rm diag}}
\def\longvert{{\rule[-2mm]{0.1mm}{7mm}}\,}
\def\a{\alpha}
\def\dag{{}^{\dagger}}
\def\tq{{\widetilde q}}
\def\p{{}^{\prime}}
\def\W{W}
\def\N{{\cal N}}
\def\hsp{,\hspace{.7cm}}

\def\br{\nonumber\\}
\def\IZ{{\mathbb Z}}
\def\IR{{\mathbb R}}
\def\IC{{\mathbb C}}
\def\IQ{{\mathbb Q}}
\def\IP{{\mathbb P}}
\def \eqn#1#2{\begin{equation}#2\label{#1}\end{equation}}

\newcommand{\C}{\ensuremath{\mathbb C}}
\newcommand{\Z}{\ensuremath{\mathbb Z}}
\newcommand{\R}{\ensuremath{\mathbb R}}
\newcommand{\rp}{\ensuremath{\mathbb {RP}}}
\newcommand{\cp}{\ensuremath{\mathbb {CP}}}
\newcommand{\vac}{\ensuremath{|0\rangle}}
\newcommand{\vact}{\ensuremath{|00\rangle}                    }
\newcommand{\oc}{\ensuremath{\overline{c}}}
\begin{titlepage}
\begin{flushright}
SISSA 69/2010/EP
%ULB-TH/09-10\\
%hep-th/yymmnnn\\
\end{flushright}
\bigskip
\def\thefootnote{\fnsymbol{footnote}}

\begin{center}
{\Large
{\bf
Type IIB Holographic Superfluid Flows
\vspace{0.1in}
%A Black Hole with Vector Hair
}
}
\end{center}

\bigskip
\begin{center}
{\large  Daniel Are\'an$^{a,b}$,
Matteo Bertolini$^{a,b}$,
Chethan Krishnan$^b$\\
\vspace{0.1in}
and Tom\'{a}\v{s} Proch\'{a}zka$^b$}

\end{center}

\renewcommand{\thefootnote}{\arabic{footnote}}

\begin{center}
%\vspace{0.2cm}
$^a$ {International Centre for Theoretical Physics (ICTP)\\
Strada Costiera 11; I 34014 Trieste, Italy \\}
\vskip 5pt
$^b$ {SISSA and INFN - Sezione di Trieste\\
Via Bonomea 265; I 34136 Trieste, Italy\\}
\vskip 5pt
{\texttt{arean,bertmat,krishnan,procht @sissa.it}}

\end{center}

\noindent
\begin{center} {\bf Abstract} \end{center}
We construct fully backreacted holographic superfluid flow solutions in a
five-dimensional theory that arises as a consistent truncation of low energy
type IIB string theory. We construct a black hole with scalar and vector hair in this theory, and study the
phase diagram. As expected, the superfluid phase ceases to exist for high enough
superfluid velocity, but we show that the phase transition between normal and superfluid phases
is always second order. We also analyze the zero temperature limit of these solutions. %that we find.
Interestingly, we find evidence that the emergent IR conformal symmetry of the zero-temperature domain
wall is broken at high enough velocity.
\vspace{1.6 cm}
\vfill

\end{titlepage}

\setcounter{footnote}{0}

%%%%%%%%%%%%%%%%%%%%%%%%%%%%%%%%%%%%%%%%%%%%%%%%%%%%%%%%%%%%%%%%%%%%%%%%%%%%%%%%%%%%%%%%%%%%%%
%%%%%%%%%%%%%%%%%%%%%%%%%%%%%%%%%%%%%%%%%%%%%%%%%%%%%%%%%%%%%%%%%%%%%%%%%%%%%%%%%%%%%%%%%%%%%%
\section{Introduction}
\label{intro}

In the last few years there has been an intense effort to model superconductor/superfluid
phase transitions using the AdS/CFT correspondence.
The basic observation that makes this industry possible is the fact that at finite charge density and
at sufficiently low temperatures, an AdS black hole in the presence of a charged scalar field
is unstable to the formation of hair \cite{Gubser:2008px}. Using the basic AdS/CFT dictionary \cite{Maldacena,GKP,Witten},
this gets easily interpreted as a superfluid-like phase transition in the dual field theory, cf.
Weinberg \cite{weinberg}.

Much of the work on holographic superconductors is done in the context of phenomenological models, along the
lines of the proposal originally presented in \cite{HHH1}. This is based on the minimal set-up of a charged massive
scalar minimally coupled to Einstein-Maxwell theory. While many interesting results can be obtained within this
minimal framework (see \cite{Hartnoll:2009sz,Herzog:2009xv,Horowitz:2010gk} for reviews and references), such a
bottom-up approach has some intrinsic limitations. Since the hope is that holographic constructions may eventually shed some
light  on some basic properties of high-$T_c$ superconductors, it would be desirable to have
a microscopic understanding of the underlying theory. This is something that phenomenological models, by definition,
cannot offer. Secondly, they do not guarantee the existence of a quantum critical point in the phase diagram, which is
instead expected to control the physics of high-$T_c$ superconductors.
Indeed, the phenomenological models that
one typically works with have no potentials but the mass term. However, it is expected that to have an emergent conformal
symmetry in the infrared in the zero temperature limit, one should have potentials that allow symmetry-breaking
minima \cite{GubRoch2}. Recently, some progress has been made in this respect and several
microscopic embeddings of holographic superconductors have been proposed
in the framework of type IIB string theory \cite{Gubser}, M-theory \cite{Gaunt}, and D7-brane models
\cite{Ammon:2009fe}. In these models, the potentials quite generically allow symmetry breaking vacua.

Most studies have also been performed in the probe
approximation, which is a large-charge limit in which the backreaction of the matter fields on
the gravitational field is negligible. While many interesting results can be obtained with such a simplified setup when the
temperatures are near the phase transition, the analysis becomes less and less
reliable at very low temperatures, where the backreaction is non-negligible. This prevents exploration of interesting
low temperature phenomena: in particular, understanding the ground state of holographic superconductors is outside the regime
of applicability of the probe limit.
Therefore, it is useful to realize holographic constructions where the backreaction is taken into account.
Progress in this direction began with \cite{HHH2}, where a (numerical) backreacted solution for the phenomenological model
of \cite{HHH1} was presented.

In trying to explore the phase diagram of holographic
superconductors, an interesting direction was pursued in
\cite{Basu,Herzog} where the original holographic superfluid was
studied in the presence of a non-vanishing superfluid
velocity (aka superfluid flow). Holographically this needs a non-trivial profile for a
spatial component of the gauge field, besides the ever-present
temporal component. The latter corresponds to a charge density  and is
necessary to have a phase transition in the first place (see
\cite{Faulkner:2010gj} for a recent  alternative proposal). Two
interesting results obtained in \cite{Basu,Herzog} were to show the
existence of a critical velocity above which the superfluid phase
ceases to exist, as expected for physical superfluids, and the
existence of a tricritical point in the velocity vs. temperature
diagram where the order of the phase transition changes from
second to first. Moreover, it was noticed in \cite{Keranen:2009re,Daniel} that these
solutions can be efficiently compared to 2+1-dimensional
superconducting thin films or wires. They behave very much like
superfluids, in that an applied external magnetic field does not get
expelled as if the gauge field were not dynamical. The
four-dimensional gravitational model of \cite{Basu,Herzog} was further
analyzed from this latter viewpoint in \cite{Daniel}, where the system
was in fact studied  at fixed current rather than at fixed velocity.
This choice allowed new checks, and remarkable agreement with some
peculiar properties of real-life superconducting films (see
\cite{Tinkham})  was found.

All solutions  presented in \cite{Basu,Herzog,Daniel} have been
obtained in the probe approximation.  Hence, while being able to
confront  phenomena near or right below the critical temperature, not
much could be said about the low temperature regime of such
superfluid flows. This problem was addressed more recently in \cite{Tisza},
where the backreaction of the phenomenological four-dimensional model of
\cite{Basu,Herzog} was obtained.

%%%%%%%%%%%%%%%%%%%%%%%%%%%%%%%%%%%%%%%%%%%%%%%%%%%%%%%%%%%%%%%%%%%%%%
\subsection{Summary of Results}
\label{Results}

In this paper we take some concrete steps forward in the above program on
superfluid flows: we focus our attention on models with known microscopic
embedding and symmetry breaking vacua, and work at the backreacted
level. Specifically, we will describe a holographic superfluid flow in four
dimensions by means of a fully backreacted solution of a five-dimensional
gravitational system whose action arises  as a consistent truncation of type IIB
string theory \cite{Gubser}. The effective theory is essentially Einstein-Maxwell theory with a Chern-Simons term,
interacting with a complex charged scalar with a non-trivial potential. It can be obtained
upon compactification of type IIB theory on an $AdS_5 \times Y$ geometry, $Y$ being a
Sasaki-Einstein manifold. Using the numerical solutions that we find, we
analyze several aspects of the rich phase diagram of
this system. In particular, we present the plots of the scalar
condensate against temperature and its dependence on the
superfluid velocity, analyze the nature of the phase transition
computing the free energy difference between the superconducting and
the normal phase, and give some predictions on the zero
temperature limit.

As one would expect on physical grounds, we observe that for high
enough velocity the system stops superconducting. Interestingly, we
find that for all velocities we have investigated the phase
transition in these type IIB constructions is always second
order. Hence, we do not find the tricritical  point which
characterizes the phase diagram of models with large charges. The same
behavior was  observed in the phenomenological but backreacted $AdS_4$
model of \cite{Tisza} for low values of  the scalar charge (in fact,
only for the case $q=1$ in their notation).  The persistence of the
second order phase transition has been observed also in the
unbackreacted case for large masses of the scalar in five dimensions
\cite{Daniel2}. We will have some more comments on this in section
\ref{phtr}.

One of the advantages of having a fully backreacted model is that one
can also investigate the low temperature limit. In the zero velocity
case, it is known \cite{GubRoch, GubRoch2} that the type IIB hairy
black hole solution tends to a domain wall with an emergent conformal
symmetry in the deep IR. (This is in contrast with the
phenomenological model of \cite{HoroRob} where the potential  has only
a mass term and no symmetry breaking minima, and the zero temperature
limit generically does not lead to an IR AdS geometry.) When the
velocity is turned on and it is high enough, we find evidence that the
solution stops being AdS in the IR. This suggests that beyond some
critical velocity the IR conformality is lost. Along the way, we also
discuss the importance of the frame comoving with the superfluid flow in these results.

The rest of the paper is organized as follows. In section \ref{5d} we
present the truncated type IIB five-dimensional action and the
equations of motion. Our ansatz for the relevant fields,  and the
procedure we pursue to obtain numerical solutions with the desired
features are discussed in section \ref{numer}.  Using these solutions,
in section \ref{phtr} we study the phase diagram of the superfluid
flow. In particular, we analyze the nature of the phase transition as
a function of the superfluid velocity. In doing so, we compute the
free energy of the superfluid phase and compare it to that of the
normal phase (which is described, holographically, by a
Reissner-Nordstrom black hole with no scalar hair). Finally, in
section \ref{zeroT} we study the $T \rightarrow 0$ limit of some
geometrical quantities like the Ricci scalar and the Riemann tensor
squared. We also study the variation %evolution 
of the superfluid fraction %with the temperature.
as the temperature is lowered. 
These analyses allow us to explore the nature of the
ground state of holographic type IIB superfluid flows.
The appendices contain more technical
material which might help the reader in following our analytical and
numerical computations more closely.

%%%%%%%%%%%%%%%%%%%%%%%%%%%%%%%%%%%%%%%%%%%%%%%%%%%%%%%%%%%%%%%%%%%%%%%%%%%%%%%%%%%%%%%%%%%%%%%%%%%%%%%%
%%%%%%%%%%%%%%%%%%%%%%%%%%%%%%%%%%%%%%%%%%%%%%%%%%%%%%%%%%%%%%%%%%%%%%%%%%%%%%%%%%%%%%%%%%%%%%%%%%%%%%%%
\section{The IIB Set Up}
\label{5d}

In \cite{Gubser} a consistent truncation of type IIB supergravity was
presented, which has the structure of an Einstein-Maxwell (plus
Chern-Simons) system in five dimensions coupled to a charged scalar
field with a non-trivial potential. The action reads \bea
\label{IIBac}
S_{IIB}&=&\int d^5x \sqrt{-g}\bigg[R-{L^2 \over 3}F_{ab}F^{ab}+ {1
\over 4}\left({2L \over 3}\right)^3 \epsilon^{abcde}F_{ab}F_{cd}A_e
+\nonumber
%\hspace{1in}
\\
%\hspace{.5in}
&&- \frac{1}{2}\left((\partial_a \psi)^2 + \sinh^2 \psi(\partial_a
\theta -2 A_a)^2-{6 \over L^2}\cosh^2\left({\psi \over
2}\right)(5-\cosh \psi)\right)\bigg]\,.  \eea Here,
$\epsilon^{01234}=1/\sqrt{-g}$, and we have written the charged
(complex) scalar by splitting the phase and the modulus in the form
$\psi e^{i\theta}$. For later convenience we recall that the Abelian
gauge field $A$ is dual to an $R$-symmetry in the boundary field
theory \cite{Gubser} and the scalar field has $R$-charge $R=2$.

The matter equations of motion are \bea &&{1 \over
\sqrt{-g}}\partial_a\Big({4 \over 3}L^2\sqrt{-g}F^{ab}-{8 \over
27}L^3\sqrt{-g}\epsilon^{abcde}F_{cd}A_e\Big)+
%\hspace{1.85in}
\nonumber \\ &&\hspace{1.55in}+{2 \over
27}L^3\epsilon^{pqrsb}F_{pq}F_{rs}+2 \sinh^2 \psi(\partial^b \theta- 2
A^b)=0\,, \\ \nonumber \\ &&{1 \over \sqrt{-g}}\partial_a
(\sqrt{-g}\partial^a \psi)-\frac{1}{2}\sinh 2 \psi (\partial_b \theta-
2 A_b)^2+
%\hspace{1.85in}
\nonumber \\ &&\hspace{1.55in}+\frac{3}{2 L^2}\left(\sinh \psi
(5-\cosh \psi)-2\cosh^2 \left({\psi \over 2}\right) \sinh
\psi\right)=0\,.  \eea The Einstein equations can be written as \bea
&&R_{ab}-\frac{1}{2}g_{ab} R -{2 \over 3} L^2\Big(F_{ac}F_b^{\
c}-\frac{g_{ab}}{4}F^{cd}F_{cd}\Big)+%\hspace{1.85in}
\nonumber \\
&&\hspace{1.55in}-\frac{1}{2} \Xi_{ab}+\frac{1}{4} g_{ab} \Xi_a^{\
a}-\frac{3}{2 L^2} g_{ab}\cosh^2\left({\psi \over 2}\right) (5-\cosh
\psi)=0\,, \\
%\nonumber \\
&&{\rm with}  \ \  \Xi_{ab} \equiv \partial_a \psi \partial_b \psi +
\sinh^2 \psi (\partial_a \theta - 2 A_a)(\partial_b \theta - 2 A_b)\,.
%\hspace{1in}
\nonumber \eea It is convenient to use the gauge invariance to shift
away the angle $\theta$ and also write the various expressions in
terms of covariant derivatives. This basically means that we set
$\theta$ to zero in the above equations and use \bea &&\nabla_a\Big({4
\over 3}L^2F^{a}_{\ b}-{8 \over 27}L^3\epsilon^{a\ cde}_{\
b}F_{cd}A_e\Big)%+\hspace{1.85in}\nonumber \\
%\hspace{1.85in}
+{2 \over 27}L^3\epsilon^{pqrs}_{\ \ \ \ b}F_{pq}F_{rs}-4 \sinh^2 \psi
\ A_b=0\,,
%\hspace{0.45in}
\label{eommatt1} \\
&&\nabla_a\nabla^a \psi-2 \sinh 2 \psi
(A_bA^b)%+\hspace{1.85in}\nonumber \\
%\hspace{1.85in}
+\frac{3}{2 L^2}\Big(\sinh \psi (5-\cosh \psi)-2\cosh^2\left({\psi
\over 2}\right) \sinh \psi\Big)=0\,,\hspace{.35in}\label{eommatt2}
\eea as the matter equations of motion. The leading terms in the
scalar potential take the form \bea V(\psi)=-{12 \over L^2}- {3 \psi^2
\over 2L^2}+...  \eea which have the immediate interpretation as the
AdS cosmological constant and the scalar mass term. Typically, in a
minimal phenomenological model the scalar potential has just the above
two terms. Higher order terms affect mostly the very low temperature
regime where the condensate becomes larger and thus the type IIB model
can become substantially different from the minimal one. There are
then two reasons as to why one should try and work out a fully
backreacted solution for this type IIB model. The first is that the
scalar has charge $R=2$ and hence the probe approximation, which is a
large charge scaling limit, is potentially inappropriate already at
temperatures near the critical temperature. The second is that a
backreacted solution would let one study the system in a regime
(i.e. very low temperatures) where, as just noticed, the differences
of the action (\ref{IIBac}) with respect to that of a phenomenological
model, are more apparent.

Note that the scalar mass is $m^2=-3$. In $d=4$, this mass is in the
range where the leading fall-off at the boundary, which is ${\cal
O}(1/r)$, corresponds to a non normalizable mode.  So, using the
AdS/CFT map, we will interpret it as the source of the dual field
theory operator ${\cal O}$.  The subleading fall-off is ${\cal
O}(1/r^3)$ and corresponds to a condensate for ${\cal O}$ (whose
dimension will therefore be $\Delta=3$). It is evident from the value
of the R-charge and  this fall-off that $\Delta=3|R|/2$ and ${\cal O}$
is therefore a chiral primary \cite{Gubser}.

%%%%%%%%%%%%%%%%%%%%%%%%%%%%%%%%%%%%%%%%%%%%%%%%%%%%%%%%%%%%%%%%%%%%%%%%%%%%%%%%%%%%%%%%%%%%%%
%%%%%%%%%%%%%%%%%%%%%%%%%%%%%%%%%%%%%%%%%%%%%%%%%%%%%%%%%%%%%%%%%%%%%%%%%%%%%%%%%%%%%%%%%%%%%%
\section{Hairy Black Hole Solution}
\label{numer}

We want to construct a fully backreacted hairy black hole solution,
holographically describing a superfluid flow. To achieve this we must
keep the metric (also) unfixed and find a self-consistent solution for
the metric, the gauge field and the scalar. To have a charged scalar
condense, we need to turn on both the scalar and the time component of
the gauge field in the bulk \cite{Gubser:2008px}.  Moreover, to obtain
a non-vanishing superfluid flow, we should break the isotropy in the
boundary directions that was present in the original holographic
superconductor construction of \cite{Gubser}. Indeed, the superfluid
velocity in (say) the $x$-direction is captured by the leading
fall-off of the bulk gauge field component $A_x$ at the boundary,
which should therefore have a non-trivial bulk profile. Altogether,
this means that we need $\psi, A_t$ and $A_x$ to be non-trivial. Since
we would like to work with ordinary as opposed to partial differential
equations, we look for an ansatz where these are functions purely of
the holographic direction $r$ : fortunately, this turns out to be
enough to obtain a solution. Consistency of the Einstein equations
then demands that we choose a metric ansatz of the form
\beq
ds^2=-{r^2 f(r) \over L^2}dt^2+{L^2 h(r)^2 \over r^2 f(r)} dr^2 -2
C(r) \frac{r^2}{L^2}dt dx+{r^2 \over L^2}B(r) dx^2+{r^2 \over L^2}
dy^2 + {r^2 \over L^2} dz^2\,.
\label{ModTisza4}
\eeq The metric contains four independent functions, $f(r),h(r),C(r)$
and $B(r)$. Together with the ansatz for the gauge field and the
scalar \bea A=A_t(r)\, dt + A_x(r) \,dx\;, \qquad \psi=\psi(r)\,,
\label{gauge3}
\eea this will give rise to a set of seven independent equations for
seven unknowns. Our ansatz here is essentially of the same form as the
one in \cite{Tisza}, albeit in one more dimension. This can be
demonstrated by going over to an Eddington-Finkelstein form and
working in a frame where the normal fluid considered in \cite{Tisza}
is at rest.

Let us first notice that with this choice of ansatz the terms in the
equations of motion (\ref{eommatt1}) arising from the
$\epsilon^{abcde}$ piece all vanish. A second important fact is that
there are several scaling symmetries one should be aware of. In
particular, the ambiguity in the units at the boundary for the time
$t$ and the distance along $x$ translate to the following two scaling
symmetries of the resulting equations \bea && t \rightarrow
t/\mbox{a}\;, \quad f \rightarrow \mbox{a}^2 f\;, \quad h\rightarrow
\mbox{a} \,h\;, \quad C \rightarrow \mbox{a} \,C\;, \quad A_t
\rightarrow \mbox{a} \,A_t\;,
\label{scale1B} \\
&& x \rightarrow x/\mbox{b}\;, \quad B \rightarrow \mbox{b}^2 B\;,
\quad C \rightarrow \mbox{b}\, C\;, \quad A_x \rightarrow \mbox{b}\,
A_x\;.\label{scale2B}
%\hspace{0.3in}
\eea These are symmetries of the action and therefore of the equations
of motion. Two further scaling symmetries of the system that we will
use are \bea\label{scale3B} && (r, t, x, y, z, L) \rightarrow \alpha
(r, t, x, y, z, L)\;, \quad (A_t, A_x) \rightarrow (A_t, A_x)/\alpha\;,
%\hspace{0.05in}
\\
\label{scale4B}
&& r \rightarrow \beta r\;, \quad (t, x, y, z) \rightarrow (t, x, y,
z)/\beta\;, \quad (A_t, A_x) \rightarrow \beta (A_t, A_x)\;.  \eea The
first scaling changes the metric by a factor $\alpha^2$ and leaves the
gauge field invariant, but its effect is to scale the action
(\ref{IIBac}) by an overall constant factor $\alpha^2$, therefore
leaving the equations of motion unaffected. The second scaling is the
usual holographic renormalization group operation in AdS, and it is
easily seen that the metric, gauge field and the equations of motion
are left invariant. Using the symmetries (\ref{scale3B}) and
(\ref{scale4B}) we can scale the horizon radius $r_H$ and the AdS
scale $L$ to unity. We will assume this has been done in what follows,
unless stated otherwise.

The strategy we pursue to construct (numerically) our solution is as
follows. First, using our ansatz, one can massage the equations of
motion and end up with first order differential equations for $f$ and
$h$ and second order differential equations for $B, C, A_t, A_x$ and
$\psi$. All in all we have then two first order and five second order
equations resulting in twelve degrees of freedom. Therefore, to fix a
solution we need twelve pieces of data.

We start by considering the fields (\ref{ModTisza4})-(\ref{gauge3})
near the horizon ($r=r_H$) and expand their several components $\Phi$
in a Taylor series as \beq
\label{expH}
\Phi = \Phi_0^H + \Phi_1^H (r - r_H) + \dots\,.  \eeq Requiring
regularity of the solution at the horizon amounts to setting some
specific coefficients to zero. To linear order in $(r-r_H)$, the
expansion at the horizon takes the form \bea
f&=&f_1^H(r-r_H)+... \label{start}\\ h&=&h_0^H+h_1^H(r-r_H)+... \\
B&=&B_0^H+B_1^H(r-r_H)+...\\ C&=&C_1^H (r-r_H)+... \\
A_t&=&A_{t,1}^{H}(r-r_H)+... \\
A_x&=&A_{x,0}^{H}+A_{x,1}^{H}(r-r_H)+...\\
\psi&=&\psi_0^H+\psi_1^H(r-r_H)+...\,.
\label{end}
\eea That is, demanding regularity is tantamount to setting $f_0^H,
C_0^H$ and $\phi_0^H$ to zero.  Imposing now the equations of motion
has the effect of putting further constraints on many coefficients,
which all end up being determined by a small set of independent
horizon data. It turns out that the coefficients can all be determined
in terms of six independent data \bea (h_0^H, B_0^H, C_1^H, A_{t,1}^H,
A_{x,0}^H, \psi_0^H)\,.
\label{horizondata5d}
\eea This means that the solutions that we will find by integrating
from the horizon will be a six-parameter family. All other
coefficients are functions of these ones.  One such relation which
will be useful later is \bea
\label{f1H}
f_1^H=(h_0^H)^2\Big({9 \over 4}+2 \cosh{\psi_0^H}-{\cosh(2\psi_0^H)
\over 4}\Big)-\frac{2 (A_{t,1}^H)^2}{9}\,.  \eea The next step is to
integrate the solution from the horizon out to the boundary
($r\rightarrow \infty$), starting with the free horizon data
(\ref{horizondata5d}), trying a suitable ansatz for the asymptotics of
the fields at the boundary. In fact, the asymptotic expansion in five
dimensions is subtle because, as already noticed, the mass of the
scalar is such that there is a non-normalizable mode. To accommodate a
generic solution obtained by integration from the horizon, we
therefore need to turn on the non-normalizable mode of the scalar as
well at the boundary. The non-normalizable mode triggers further
logarithmic terms in the asymptotic expansion, so we need to keep
track of them as well. It turns out that a combined series expansion
in both $1/r^n$ and $\log r/r^m$ \bea \Phi = \sum_{n=0}^{\infty}
\Phi_n\,\frac{1}{r^n} + \sum_{m=0}^{\infty} \Phi^l_m \,\frac{\log
r}{r^{m}}
\label{boundexp5d}
\,, \eea works nicely.

Using a shooting technique we select, out of all possible solutions,
those which match our physical requirements. In particular, we ask
that the space be asymptotically AdS and that the source term for the
field theory operator dual to the scalar field be vanishing, since we
want the $U(1)$ breaking to be spontaneous.

We have found that the following asymptotic expansion solves the
equations of motion\footnote{What we do is to plug this expansion into
the EoMs and demand that the result be zero order-by-order. We find
that either this is satisfied identically or that the resulting
relations can be interpreted as the definitions of higher order terms
in the expansion.}, while being general enough to match the curves
arising from the integration from the horizon \bea && f=h_0^2+
\frac{f_4}{r^4}+\frac{f^l_4}{r^4} \log r+...\;, \qquad\quad\;\;
h=h_0+{h_2 \over r^2}+{h_4 \over r^4}+{h^l_4 \over r^4}\log r+...\;,
\label{falloff1}\\
&& B=B_0+\frac{B_4}{r^4}+\frac{B^l_4}{r^4}\log r+...\;, \qquad
\hspace{-0.15in} 
\quad\;\; C=C_0+\frac{C_4}{r^4}+\frac{C^l_4}{r^4}\log
r+...\;, \label{falloff2}\\
&& A_t=A_{t,0}+\frac{A_{t,2}}{r^2}+\frac{A^l_{t,2}}{r^2}\log
r+...\;, \quad A_x=A_{x,0}+{A_{x,2} \over
r^2}+\frac{A^l_{x,2}}{r^2}\log r+...\;,\label{falloff3}\\
&&\psi=\frac{\psi_1}{r}+\frac{\psi_3}{r^3}+\frac{\psi^l_3}{r^3}\log
r+...\;.\label{falloff4}
\eea
Of course, not all of the above coefficients are independent. We
relegate the explicit expressions for the dependent ones to Appendix
\ref{asymcoef}. We merely note that when the non normalizable mode
$\psi_1$ is set to zero, the expressions are such that all the
logarithmic pieces vanish as expected. It can also be seen that the
independent parameters at the boundary can be taken to be \beq (h_0,
f_4, B_0, B_4, C_0, C_4, A_{t,0}, A_{t,2}, A_{x,0}, A_{x,2}, \psi_1,
\psi_3)\,.  \eeq To get asymptotically AdS solutions, we must set
$B_0, h_0$ to 1 and $C_0, \psi_1$ to zero. The scaling symmetries can
be used to accomplish the first two conditions, whereas we need to
shoot for the last two. We are therefore left with eight independent
boundary data. They are \beq (f_4, B_4, C_4, A_{t,0}, A_{t,2},
A_{x,0}, A_{x,2}, \psi_3)\,.  \eeq We see here that the physical
requirements we impose at the boundary do not fix as many integration
constants of the ODE system, as the regularity conditions at the
horizon does. This means that for the solutions that we obtain, there
are hidden relations between the boundary data. Concretely, since
there are only six independent pieces of horizon data this gives us
relations between the above eight variables, which will then be used
to study the phase diagram of the boundary theory.

An important quantity for studying the thermodynamics of the system is
of course the superfluid temperature. This corresponds to the black
hole Hawking temperature, T. %Since the black hole is stationary but not
%static, the determination of the Hawking temperature via Wick rotation is formally 
%not correct in this case. This is because Wick-rotating time does not result in a real metric for such
%black holes.  Instead we will compute the surface gravity $\kappa$
%directly, which is related to the Hawking temperature as $T =
%\kappa/2\pi$. 
From the structure of the metric (\ref{ModTisza4}) we
easily get 
\beq
\label{temp4d}
T = \frac{r_H^2\, f'(r_H)}{4\pi\,L^2\,h(r_H)}\,, 
\eeq 
which is then
also determined in terms of our horizon data. After some
simple algebra, recalling we have set $r_H=1$ and using the horizon
relation (\ref{f1H}), we get
\beq
T = \frac{1}{4\pi}\left[h_0^H
\left({9 \over 4}+2 \cosh{\psi_0^H}-{\cosh(2\psi_0^H) \over 4}\right)-
\frac{2 (A_{t,1}^H)^2}{9 h_0^H} \right]
\label{temp5d}\,.
\eeq

%%%%%%%%%%%%%%%%%%%%%%%%%%%%%%%%%%%%%%%%%%%%%%%%%%%%%%%%%%%%%%%%%%%%%%%%%%%%%%%%%%%%%%%%%%%%%%%%%%
%%%%%%%%%%%%%%%%%%%%%%%%%%%%%%%%%%%%%%%%%%%%%%%%%%%%%%%%%%%%%%%%%%%%%%%%%%%%%%%%%%%%%%%%%%%%%%%%%%
\section{Superfluid Flow Phase Transition}
\label{phtr}

We plot the result for the condensate versus the temperature in figure 1, for different values of the
superfluid velocity. For the rest of the paper, we will introduce the notation
\beq
\mu \equiv A_{t,0}, \ \ {\langle {\cal O}\rangle}\equiv{\sqrt{2}\,\psi_3}, \ \ \xi \equiv \frac{A_{x,0}}{A_{t,0}}\,,
\eeq
where $\mu$ is the field theory chemical potential, ${\cal O}$ the (condensing) chiral primary operator, and $\xi$
the superfluid velocity in units of the %black hole 
chemical potential. When we work in an ensemble with fixed  chemical
potential, the meaningful (dimensionless) quantities relevant for the condensate plot are
\beq
\frac{T}{\mu} \ \ {\rm and} \ \ \frac{\langle {\cal O}\rangle}{\mu^3}\,.
\eeq
In constructing the plots, we have also rescaled by the (velocity-dependent) factor $\sqrt{1-\xi^2}$, which is
nothing but the relativistic boost factor.

From the form of the curves in figure 1, it is evident that there is a
phase transition to a hairy black hole at low temperatures.
\begin{figure}[ht]
\begin{center}
\includegraphics[height=0.3\textheight]{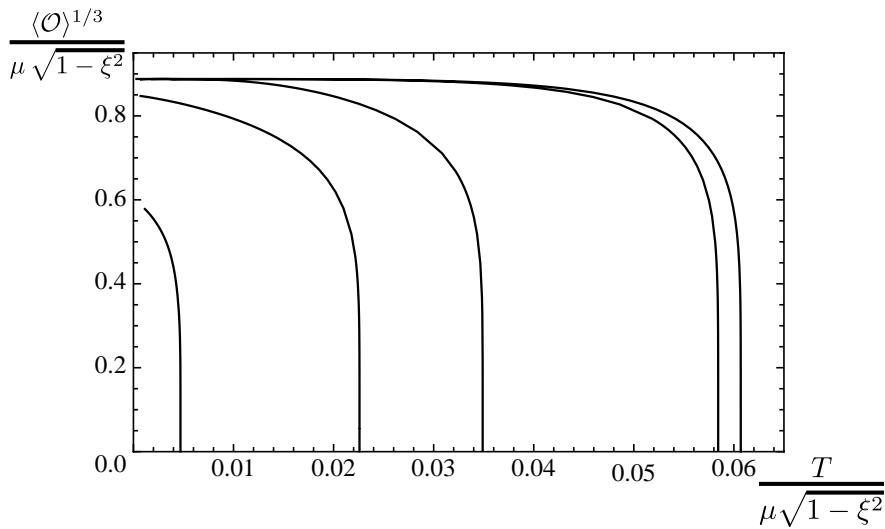}
\label{condplot}
\vskip -10pt
\caption{Condensate plots for various values of the velocity
$\xi=0,0.1,0.33,0.4,0.5$ (from right to left). The zero velocity case,
$\xi=0$, which we report for ease of comparison, precisely agrees with
existing results in the literature \cite{Gubser}.}
\end{center}
\end{figure}
As expected, the critical temperature decreases as the velocity is
increased. For instance, for $\xi=1/2$ (which is the highest velocity
we have investigated) we observe that
$T_c(\xi=1/2)=0.067\,T_c(\xi=0)$. It is clear from the condensate plot
that the superfluid  phase cannot exist for velocities that are much
higher than this.
\begin{figure}[ht]
\begin{center}
\includegraphics[height=0.3\textheight]{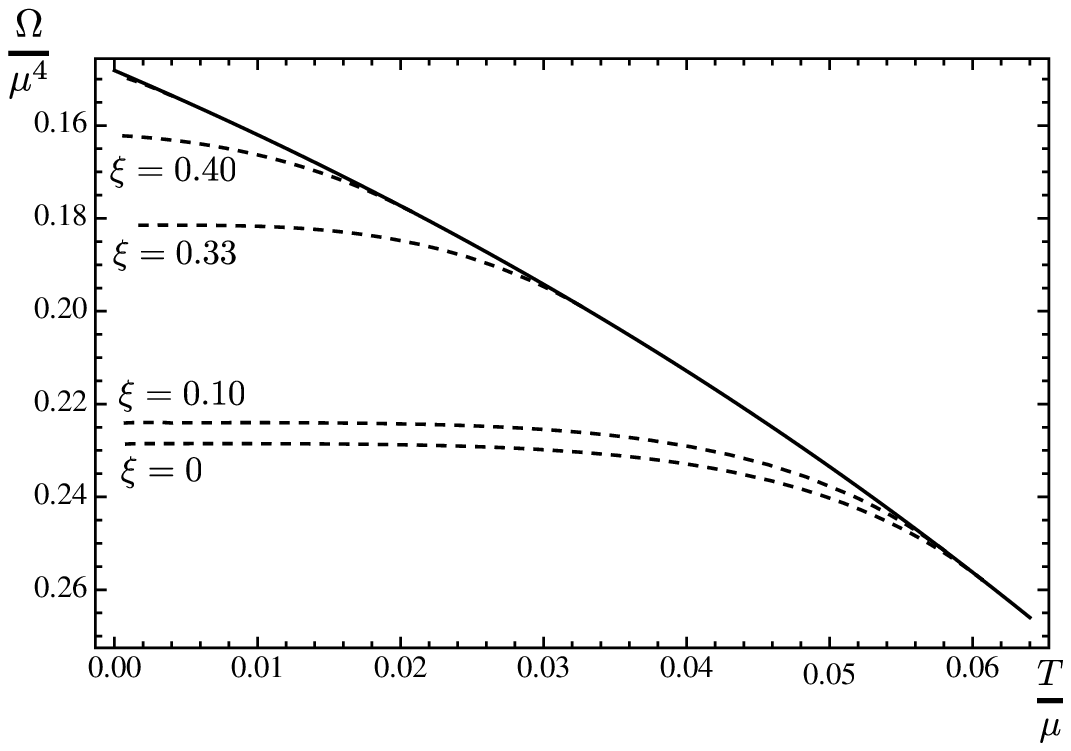}\label{freEplot}
\vskip -10pt
\caption{Free energy plots for various velocities $\xi=0, 0.1, 0.33,
0.4, 0.5$ (dashed lines from bottom to top). The RN-AdS black hole is
also presented for comparison (solid line). The plots show that for
any velocities the phase transition is second order. The apparent overlap of the
$\xi=0.5$ curve with the normal phase, is an artifact of the resolution of
the figure.}
\end{center}
\end{figure}

One can compare the free energy of the normal phase (which corresponds
to a Reissner-Nordstrom black hole with no  hair) and the
hairy/superfluid phase to see that the superfluid phase is favored
when it exists. We collect some details of the free energy computation
in appendix \ref{RN}, while figure 2 contains the free energy
comparison between the superfluid phase and the normal phase at the
same value of  $T/\mu$. In terms of $S_{\rm ren}$  defined in
eq.~(\ref{Sren}), the precise quantities we plot are \beq \frac{S_{\rm
ren}}{\mu^4\, {\rm Vol}_4}\equiv \frac{\Omega}{\mu^4}\quad {\rm vs.}
\quad \frac{T}{\mu}\,.  \eeq The plot demonstrates that the phase
transition stays second order for all values of the velocity, up to
our numerical precision. This should be contrasted to the
unbackreacted cases previously considered in the literature, where the
phase transition typically changes to first order for high enough
values of the velocity \cite{Basu,Herzog,Daniel}.  In \cite{Tisza} a
backreacted superfluid in $AdS_4$ was considered and it was found that
for low enough values of the charge of the scalar field, the phase
transition remained second order. Our type IIB system seems to be
analogous to this latter scenario: the (R-)charge of the scalar in our
case is fixed by the IIB construction to be 2 and it is plausible that
this is  a low enough value so that the transition remains second
order all through.

In \cite{Daniel2}, the phases of the (unbackreacted) superfluid for
various values of the masses of the scalar field in $AdS_5$ were
investigated and it was found that for high enough mass, there is
always a second order transition close to the normal phase. Since the
probe limit is a large charge limit, we should expect a similar
structure also in the backreacted case when the charge is large. That
is, when the charge and the mass are both large, we should expect a
persistent second order transition. In our IIB case, we are exploring
the opposite limit, namely low (R-)charge and low mass (since the
charge and mass are related for chiral primaries). Again, we find that
the second order transition exists irrespective of the velocity.
Based on these observations, it is tempting to make the suggestion
that whenever the mass and charge are scaled together in some
appropriate way, the second order transition persists for all
velocities. Of course, to make and/or establish a precise statement
along these lines will require a much more thorough exploration of the
masses and charges of the scalars than we have undertaken
here. Moreover, as already noticed, the persistence of the second order transition was
also found in the $AdS_4$ case for small charges and small mass
\cite{Tisza}, while it was found not to exist for any value of the
mass in the probe limit \cite{Daniel2}. So it is clear that the
appropriate statement, if it exists, will have to be
dimension-dependent.

%%%%%%%%%%%%%%%%%%%%%%%%%%%%%%%%%%%%%%%%%%%%%%%%%%%%%%%%%%%%%%%%%%%%%%%%%%
%%%%%%%%%%%%%%%%%%%%%%%%%%%%%%%%%%%%%%%%%%%%%%%%%%%%%%%%%%%%%%%%%%%%%%%%%%
\section{Zero Temperature Limit}
\label{zeroT}

One of the advantages of having a fully backreacted solution is that one can reliably
go to the zero temperature limit. At zero velocity, the zero temperature solution is
expected to be described by a domain wall, corresponding to the symmetry-breaking
vacuum of the scalar potential that restores conformal symmetry in the IR. Such domain wall solution
was constructed in \cite{GubRoch}, and conjectured to correspond to the ground state of the type
IIB holographic superconductor. Since we have here fully backreacted solutions at non-zero velocity, a
natural question one would like to answer is whether and how such IR behavior gets modified when the
superfluid flows.

As a warm-up, and for later comparison, let us first consider the static case. A preliminary check one can perform is to
see whether for $\xi=0$ our condensate value tends to the condensate value found in
\cite{GubRoch}. This is indeed the case: for the lowest temperature point
($T/\mu = 3.05 \cdot 10^{-4}$), our condensate in the normalizations of \cite{GubRoch}
\bea
\langle {\cal O} \rangle_{DW}\equiv \frac{\psi_3}{(2 \mu/\sqrt{3})^3}
\eea
%we find
is $\approx 0.3215$, which is close enough to the zero-temperature value of $\approx 0.322$
found in \cite{GubRoch}.

Even without explicitly constructing the domain wall solution, one can find evidence for its
existence by investigating the horizon values of the curvature scalars $R$ and $R_{abcd}R^{abcd}$.
This strategy was adopted in \cite{Gaunt} for superconductors in M-theory, and it was found that
these curvature scalars on the horizon go to the $AdS_4$ values expected from a domain wall solution
with a symmetry-breaking minimum in the IR. We can do the same computation here, and we do find
evidence that the solution has an emergent $AdS_5$ in the IR with the correct length scale.
Note that the IR AdS scale, as determined by the symmetry-breaking vacuum \cite{GubRoch} is
$L'=\frac{2^{3/2}}{3}$ where we have set $L=1$ in the UV. Using the fact that the Ricci scalar
for $AdS_5$ is $-20/L^2$, we find that the predicted value is $-22.5$ in the IR. A similar
computation using the $R_{abcd}R^{abcd}=40/L^4$ shows that in the zero temperature limit
we should get the value $50.625$. We plot the results for both curvature scalars in figure 3. %\ref{Ricciv0}
The plots clearly demonstrate that at low temperatures the curvatures indeed stabilize to the expected domain wall
values in the infrared. %\ref{curvscal0}
\begin{figure}
\begin{center}
\includegraphics[height=0.25\textheight ]{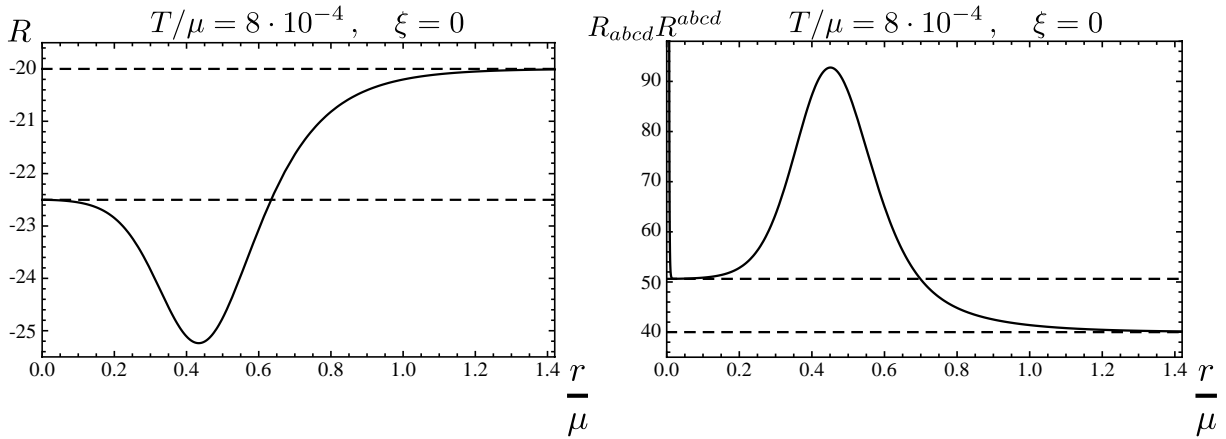}
\label{Ricciv0}
\vskip -10pt
\caption{Ricci scalar $R$ and $R_{abcd}R^{abcd}$ as a function of the radial coordinate near the horizon, at zero
superfluid velocity. The horizontal dashed lines mark the corresponding values of $R$ and  $R_{abcd}R^{abcd}$
for the UV and IR AdS geometries.}
\end{center}
\end{figure}

The behavior of $R_{abcd}R^{abcd}$ deserves a closer look, however. A distinctive feature of the present
five-dimensional case, as compared to the four-dimensional model of \cite{Gaunt}, is that $R_{abcd}R^{abcd}$
stabilizes to the domain wall value close to the horizon, but it starts  increasing %blowing up 
as the radius is further
reduced. At the horizon its value is (of course) finite, but is well on its way to the divergence at the singularity
inside the horizon\footnote{This
sharp ascent in the curvature scalars close to the horizon is not a peculiarity of the broken phase: it is also there
in the normal phase. For instance, 
$R_{abcd}R^{abcd}$, whose expression for the normal phase Reissner-Nordstorm black hole we report in eq.~(\ref{riemform}), has a similar sharp ascent at the horizon, while remaining finite there. 
%We emphasize also that this is not a numerical or coordinate artifact because outside the horizon the Schwarzschild like system is well-defined and one can get (essentially) arbitraily close to the horizon even numerically. 
On the other hand, the $AdS_4$ case is somewhat special in that the Ricci scalar is a constant
in the normal phase due to the tracelessness of the electromagnetic stress tensor in four
dimensions.}.
Note that in order to make the connection with the domain wall, what we really need is the emergence of an $AdS_5$
throat of the correct length scale at {\it zero} temperature, and our plots give evidence for that. Figure
4 %\ref{RiemannT}
reports the behavior of $R_{abcd}R^{abcd}$ zooming in near the horizon region for different temperatures. Happily, as the
temperature is lowered the stabilized region of the plot gets closer to the horizon and asymptotes to the expected
$AdS_5$ value of 50.625.
\begin{figure} [ht]
\begin{center}
\includegraphics[height=0.28\textheight ]{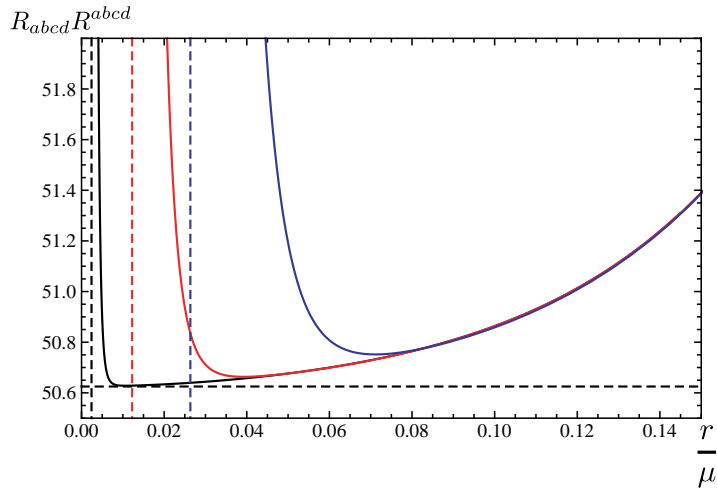}
\label{RiemannT}
\vskip -10pt
\caption{Behavior of $R_{abcd}R^{abcd}$ near the horizon for different temperatures, from left to right:
$T/\mu = 3.05 \cdot10^{-4}$ (black), $1.55\cdot10^{-3}$ (red), $3.33\cdot10^{-3}$ (blue). The
dashed vertical lines correspond to the corresponding horizon radii. The stabilized (i.e. domain wall)
value $R_{abcd}R^{abcd}=50.625$ is indicated by the dashed horizontal line.}
\end{center}
\end{figure}

Let us now consider the cases with velocity, $\xi \not =0$. We
report in figure 5 %\ref{Ricciplot}
the plot for the Ricci scalar vs. radius for different superfluid velocities (including the zero-velocity case, to
ease the comparison) and in figure 6 %\ref{Riemannplot}�
that for $R_{abcd}R^{abcd}$.  The presence
\begin{figure}
\begin{center}
\includegraphics[height=0.35\textheight ]{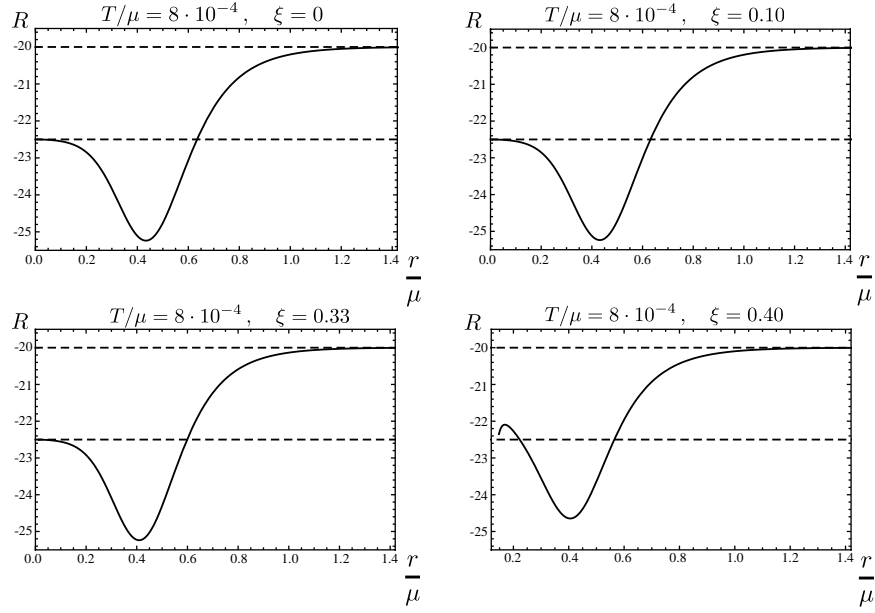}
\label{Ricciplot}
\vskip -10pt
\caption{Ricci scalar $R$ as a function of the radial coordinate near the horizon for a low temperature. The horizontal
dashed lines indicate the corresponding values of $R$ for the UV and IR AdS geometries.}
\end{center}
\end{figure}
\begin{figure}
\begin{center}
\includegraphics[height=0.35\textheight ]{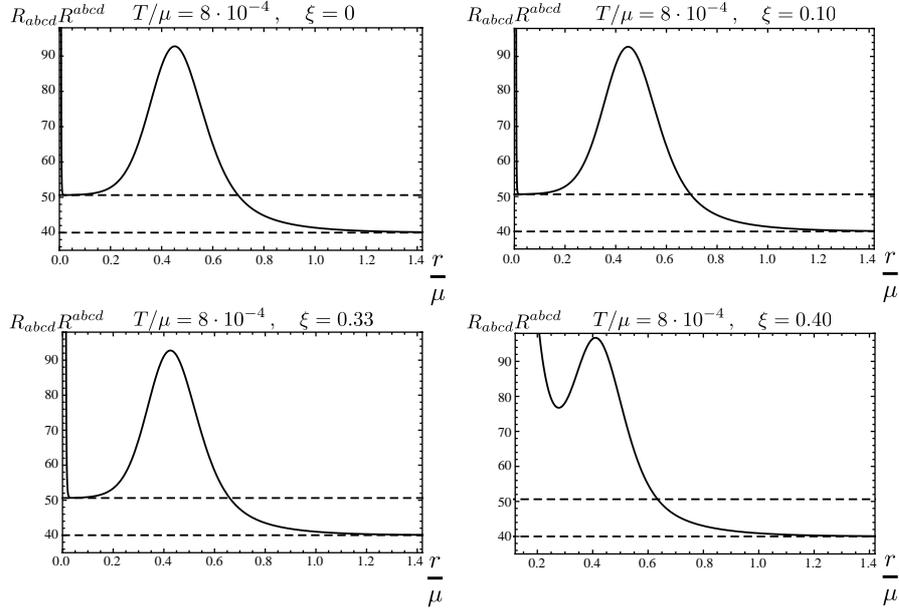}
\label{Riemannplot}
\vskip -10pt
\caption{$R_{abcd}R^{abcd}$ as a function of the radial coordinate near the horizon for a low
temperature. The horizontal dashed lines indicate the corresponding values of $R_{abcd}R^{abcd}$ for the UV and IR AdS geometries.
The stabilization to the IR value, when it happens, holds till very close to the horizon.}
\end{center}
\end{figure}
of a new scale means that there is a possibility that the emergent conformal symmetry in the IR is broken.
While for low velocities our plots suggest that the same IR fixed point as the static case is recovered,
interestingly enough, we find that for high enough velocities the conformal symmetry of the solution
is indeed broken and the curvature scalars diverge without any stabilization whatsoever. This is analogous to the
phenomenological models with no quantum critical point in the IR. The conclusion seems to be that the solutions do
not stabilize to the conformal quantum critical point when the velocity is high enough.

While we have not performed an exhaustive scan of velocities in this paper, it would be interesting to
see for what precise value of the velocity this qualitative change happens, and study the precise nature of
the phases and phase transitions, if any, there. From our analysis, it appears that the regime where this transition happens is
between $\xi=0.33$ and $\xi=0.4$. Related to this is the observation that the condensate
\bea
\frac{\langle {\cal O}\rangle^{1/3}}{\mu \sqrt{1-\xi^2}} \label{newcon}
\eea
that we plotted earlier, tends to the same value at the horizon for all values of the velocity,
for small enough velocity. This is again indicative of a quantum phase transition: there is a change
in the nature of the solution as we tune an order parameter at zero temperature. The results we find
are consistent with the idea that the phase structure in the temperature-velocity plane is determined
by the quantum critical point. It is intriguing that the relevant condensate seems to be measured in units of
chemical potential as seen in a frame comoving with the superfluid flow. For a timelike vector, which for us is the superfluid 
velocity 4-vector, the time component in the rest (i.e., comoving) frame is nothing but its norm. Therefore, since we want to 
plot a scalar quantity for the dimensionless condensate, this is the natural choice. But unlike in the case of an ordinary
fluid where the fluid velocity can be interpreted as arising from a boost of a static black hole, here the
anisotropic part of the metric does not seem to have such a simple interpretation in the bulk. We intend to come back to
some of these questions in the near future.

\begin{figure}
\begin{center}
\includegraphics[height=0.28\textheight ]{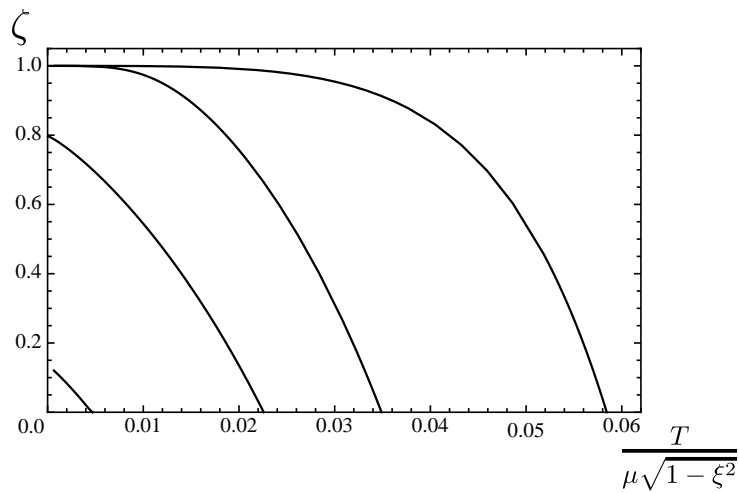}
\label{Sufracfig}
\vskip -10pt
\caption{Plots of the superfluid fraction vs. temperature for various values of the velocity 
$\xi=0.1,\;0.33,\; 0.4,\; 0.5$ (from right to left).}
\end{center}
\end{figure}
Another quantity of interest\footnote{We thank Julian Sonner for raising this point.} in understanding the 
zero temperature limit is the superfluid fraction $\zeta$. 
It corresponds to the ratio between the charge density of the superfluid flow and the total charge density of the system. 
In appendix \ref{Sfraction}, following \cite{Tisza}, we elaborate on the interpretation of the boundary theory in terms 
of a two-fluid model
and compute the expression of the superfluid fraction in terms of the fall-offs of the
bulk fields, eqs. (\ref{falloff1})-(\ref{falloff4}). The result is
\beq
\zeta=-\frac{A_{x,2}\,C_4}{A_{t,2}\,B_4}\,.
\eeq
This quantity is interesting because from the curves in figures 1 and 2 of \cite{Tisza} we see that 
for the $AdS_4$ case its behavior near zero temperature captures some interesting aspects of the nature of the phase transitions. 
More specifically, together with our results in this paper (see figure 7), we are lead to conjecture that $\zeta \rightarrow 1$ 
at zero temperature for all velocities where the rescaled condensate value at zero temperature tends to its value at zero velocity.
From the evidence presented in \cite{Tisza} one could think that %there 
the zero temperature limit of the  superfluid fraction is correlated with the existence or not of a first order phase transition 
at high enough velocity. 
However, in our case we have an explicit situation where we see a consistently second order phase transition where 
the limiting value of the condensate at zero temperature changes qualitatively as we tune the velocity. Remarkably, 
we find that $\zeta \rightarrow 1$, only in those cases where the zero temperature condensate value 
$\frac{\langle {\cal O}\rangle^{1/3}}{\mu \sqrt{1-\xi^2}}$ takes its corresponding value at zero velocity. Since 
this condensate value captures the existence or not of the (anisotropic) domain wall, the natural conjecture is that 
$\zeta=1$, for the domain wall when it exists. Notice that $\zeta \rightarrow 1$ is what one would expect for the ground 
state of a superfluid flow. What we have basically demonstrated then is that three quantities (namely the curvature 
scalar(s), the rescaled condensate and the superfluid fraction) undergo a qualitative change at the same velocity, as 
we tune the velocity. We believe this is strong evidence for the existence/non-existence of the domain wall as we go 
through that velocity.

Despite the evidence we have presented, it should be borne in mind that the preservation of the conformal
symmetry for low velocities is not fully established. Unlike in the zero velocity domain wall examples
discussed in the literature, we have not constructed an explicit solution that has emergent conformal symmetry in
the IR in the cases with (low) velocity. However, the fact that the curvature scalars and the condensate
(\ref{newcon}) stabilize to their respective zero velocity values (within our numerical precision), is an
indication that this might indeed be the case. One another caveat that we emphasize here is that the
perturbative stability of these consistent truncations in the zero temperature limit is not settled. In particular,
when the Sasaki-Einstein manifold is a sphere, instabilities are known to exist in the zero temperature
domain wall solution \cite{Girardello:1999bd,Distler:1999tr}\footnote{A related instability was recently
shown to exist also in M-theory \cite{Bobev:2010ib} for a similar consistent truncation
for the ground state of a 2+1 dimensional superconducting system \cite{GubRoch,Gauntlett:2009bh}.}. It
is possible that for a more complicated choice of Sasaki-Einstein space (which is indeed
what we need to have here anyway, in order to let the scalar chiral primary we focus on to be the
operator responsible for the black hole phase transition \cite{Gubser}) the five-dimensional theory
is stable. It is also interesting that the simple stringy consistent truncations do give rise to
scalar potentials with symmetry breaking vacua, resulting in an emergent conformal symmetry in the
IR at zero temperature. This is precisely what one expects in the zero temperature limit of a high-$T_c$
superconductor, which is believed to be governed by a quantum critical point. So our expectation is that
in the (unlikely?) event that no Sasaki-Einstein truncation can be made stable, these models should still
capture some generic features of a holographic superfluid with emergent conformal symmetry in the
IR\footnote{We thank Nikolay Bobev and Chris Herzog for a discussion on this point.}.

%%%%%%%%%%%%%%%%%%%%%%%%%%%%%%%%%%%%%%%%%%%%%%%%%%%%%%%%%%%%%%%%%%%%%%%%%%%%%%%%%%%%%%%%%%%%%%%

\section*{Acknowledgments}
\vskip -5pt
We would like to thank Silviu Pufu and Julian Sonner for email correspondence and useful comments at different stages
of this work. We also acknowledge helpful discussions and/or correspondence with Nikolay Bobev, Jarah Evslin,
Jerome Gauntlett, Chris Herzog, Giuseppe Policastro and Ho-Ung Yee. D.A.,
M.B. and C.K. would like to thank the organizers of the ESI Programme on AdS Holography and the Quark-Gluon Plasma in
Vienna, where part of this work has been done, for hospitality and financial support. C.K. thanks the International
Solvay Institutes, Brussels for hospitality during parts of this work. D.A. thanks the FRont Of Galician Speaking
scientists for unconditional support.

%%%%%%%%%%%%%%%%%%%%%%%%%%%%%%%%%%%%%%%%%%%%%%%%%%%%%%%%%%%%%%%%%%%%%%%%%%%%%%%%
%%%%%%%%%%%%%%%%%%%%%%%%%%%%%%%%%%%%%%%%%%%%%%%%%%%%%%%%%%%%%%%%%%%%%%%%%%%%%%%%

\appendix
%%%%%%%%%%%%%%%%%%%%%%%%%%%%%%%%%%%%%%%%%%%%%%%%%%%%%%%%%%%%%%%%%%%%%%%%%%%%%%%%%%%%%%%%%%%%%%%%%%%%%%
\section{Asymptotic Relations}
\label{asymcoef}

In this appendix, we present the relations defining the dependent coefficients in the asymptotic expansion in
the IIB case
\bea
&&f_4=\frac{1}{48 C_0^2}\Big(96 C_0 C_4 h_0^2+48 B_4 h_0^4+96 C_0^2 h_0 h_4+96 B_0 h_0^3 h_4+C_0^2 h_0^2 \psi_1^4+
B_0 h_0^4 \psi_1^4+
%\hspace{0.55in}
\nonumber \\
&&\hspace{0.2in}+24 C_0^2 h_0^2 \psi_1 \psi_3 +24 B_0 h_0^4 \psi_1 \psi_3
-12 h_0^4 L^4 \psi_1^2 A_{x,0}^2+24 C_0 h_0^2 L^4 \psi_1^2 A_{x,0} A_{t,0}+12 B_0 h_0^2 L^4 \psi_1^2 A_{t,0}^2\Big)\,,\nonumber\\
%\hspace{0.1in}
\\
&&f_4^l=\frac{1}{3 C_0^2}\Big(-C_0^2 h_0^2 \psi_1^4-B_0 h_0^4 \psi_1^4+3 B_0 h_0^2 L^4 \psi_1^2 A_{t,0}^2+
%\hspace{2.35in}
\nonumber \\
&&\hspace{0.2in}+C_0^2 (h_0^2 \psi_1^4-3 L^4 \psi_1^2 A_{t,0}^2)+B_0 h_0^2 (h_0^2 \psi_1^4-3 L^4 \psi_1^2 A_{t,0}^2)\Big),
%\hspace{0.1in}
\\
&& h_2=-\frac{h_0 \psi_1^2}{12}\,,\quad
h_4^l=\frac{h_0^2 \psi_1^4-3 L^4 \psi_1^2 A_{t,0}^2}{6 h_0}\,,\quad
B_4^l=L^4 \psi_1^2 A_{x,0}^2\,,\quad
C_4^l= -L^4 \psi_1^2 A_{x,0} \psi_0\,,
%\hspace{0.25in}
\\
&&\psi_3^l=-\frac{2 (C_0^2 \psi_1^3+B_0 h_0^2 \psi_1^3+3 h_0^2 L^4 \psi_1 A_{x,0}^2-6 C_0 L^4 \psi_1 A_{x,0}
A_{t,0}-3 B_0 L^4 \psi_1 A_{t,0}^2)}{3 (C_0^2+B_0 h_0^2) }\,,
%\hspace{0.5in}
\\
&&A_{t,2}^l=-\frac{3 \psi_1^2 A_{t,0}}{2}\,,\quad A_{x,2}^l=-\frac{3 \psi_1^2 A_{x,0}}{2}\,.
%\hspace{0.2in}
\eea
These are the general expressions when $\psi_1\neq0$. Our primary interest will be to shoot for
the case $\psi_1=0$, in which case all of the coefficients above vanish identically, except
\beq
h_4=\frac{f_4C_0^2-2 C_0 C_4 h_0^2- B_4 h_0^4}{2h_0( C_0^2+ B_0 h_0^2)}\,.
\eeq
In particular, all the logarithmic terms vanish and we end up with a usual asymptotic expansion in $1/r$,
as expected. Note also that in asymptotically AdS solutions, $C_0=0$ as well. Moreover, when there is no superfluid
velocity and the isotropy is not broken, $B_4=0$ and therefore we end up getting $h_4=0$. This last result is useful
in making comparisons with the holographic superconductor case investigated in \cite{Gubser}.

%%%%%%%%%%%%%%%%%%%%%%%%%%%%%%%%%%%%%%%%%%%%%%%%%%%%%%%%%%%%%%%%%%%%%%%%%%%%%%%%%%%%%%
\section{On-Shell Action and Counter-Terms}
\label{appfe}

In order to compute the free energy, we need the on-shell action for the type IIB system. As we show below it turns out
that, remarkably, the on-shell action can be written purely as a boundary piece, and be easily evaluated. However, this
boundary term is divergent: to cancel it we need to introduce boundary counter-terms. In what follows, we describe both
these steps.

For the ansatz that we work with, it can be checked directly that, despite the complications of the equations of motion,
the following relations hold
\beq
{\cal L}_0-R=\frac{2 L^2}{r^2} T_{yy} = \frac{2 L^2}{r^2} T_{zz}\,.
\eeq
Here $T$ stands for the stress tensor arising from our IIB Lagrangian, ${\cal L}_0$ is defined via
\beq
S_{IIB}=\int d^5 x \sqrt{-g} {\cal L}_0\,,
\eeq
and $R$ is the Ricci scalar. Notice that these relations only depend on our ansatz, i.e. they are true before we
use the equations of motion. Going on-shell, we replace $T_{yy}$ and $T_{zz}$ by $E_{yy}$ and $E_{zz}$, where $E$
denotes the Einstein tensor $E_{ab}\equiv R_{ab}-\frac{1}{2}g_{ab}R$. Together with the relation
\beq
E^a_{\ a}=-\frac{3}{2} R
\eeq
that is valid in five dimensions, this implies that
\beq
\sqrt{-g}{\cal L}_0=\sqrt{-g}\left(\frac{L^2}{r^2}(E_{yy}+E_{zz})-\frac 23 E^a_{\ a}\right)\,.
\eeq
The right-hand-side depends only on the metric functions and can be evaluated explicitly  for our ansatz.
Direct computation reveals that it can be written as a total differential so that the (on-shell) action takes the form
\bea
S_{IIB,{\rm OS}}= -{\rm vol}_4 \int_{r_H}^{\infty} dr \left(\frac{2 r f(r)}{L^2 h(r)^2 } \sqrt{-g}\right)'\,,
\ \ {\rm where} \ \ \sqrt{-g}= \frac{r^3h(r)}{L^3} \sqrt{\frac{C(r)^2}{f(r)}+B(r)}\,,
\eea
and the prime denotes the derivative with respect to $r$. Because of the presence of $f$, this expression is
zero at the horizon and that end of the integral is  safe.
But it clearly gets contributions from the boundary, where it diverges as $r^4$ and we need to regulate it with
appropriate counter-terms.

The counter-terms\footnote{We loosely refer to the Gibbons-Hawking term also as a counter-term, even though
strictly speaking it is a boundary term necessary to make the variational problem well-defined.} for the
gravitational part of the action in asymptotically AdS spaces can be looked up in \cite{Bala}. Along with these we also
have to add counter-terms for the scalar part. The final form of these terms in our notations and conventions
can be written as
\beq
S_{\rm ct}=2 \int d^4x \sqrt{-\gamma}\Big({\cal K}-\frac{3}{L} \Big)+\int d^4 x \sqrt{-\gamma} \frac{|\psi|^2}{L}\,.
\label{BalCT}
\eeq
The sign convention for the extrinsic  curvature is chosen so that with the outward pointing normal $n^a$,
\beq
K_{ab}\equiv \frac{1}{2}(\nabla_a n_b + \nabla_b n_a)\,.
\eeq
Note that the general gravitational counter-term discussed in \cite{Bala} involves a boundary Ricci scalar as well:
but this does not contribute for us, because our boundary becomes flat as we take it to infinity. The various
quantities (including the scalar extrinsic curvature) can be computed by cutting off the spacetime at some finite
$r=r_0$, then taking the limit $r_0\rightarrow \infty$  for the quantity $S_{IIB,{\rm OS}}+S_{\rm ct}$ at the end of
the computation. If we define the boundary at $r=r_0$, then the outward normal to the surface $\Phi(t,r,x,y,z)
\equiv r-r_0=0$ is $n_a \sim \nabla_a\Phi$, and after normalizing\footnote{Note that the boundary is timelike,
so it has a spacelike normal.} so that $g^{ab}n_an_b=1$, we get
\beq
n_a=\left(0,\frac{L h(r)}{r \sqrt{f(r)}},0,0,0\right)\,.
\eeq
Since we need only the scalar extrinsic curvature, we don't need to introduce 4-D coordinates on the boundary
and can compute it directly in the bulk coordinates as
\beq
K=g^{ab}\nabla_a n_b = \frac{ f^{1/2} \left(8 C^2+r f B'+2 r C C'+8 B f+r B f'\right)}{2 L (C^2+B
f)h}\,.
\eeq
So the final form of the counter-term action is
\bea
S_{\rm ct}= {\rm Vol}_4 \ \lim_{r \rightarrow \infty}\left[ \frac{ r^4 f^{1/2}
\left(8 C^2+r f B'+2 r C C'+8 B f+r B f'\right)}{ L^5 h \sqrt{C^2+B f}}  %\hspace{0.5in}\nonumber \\
-\frac{r^4}{L^4}\sqrt{C^2+B f} \Big(\frac{6}{L}- \frac{\psi^2}{L}\Big) \right]\,.\nonumber \\
\eea
With the addition of this piece, the renormalized action $S_{IIB,{\rm OS}}+S_{\rm ct}$ no longer has the $r^4$
divergence and is finite. The net result is
\bea
S_{\rm ren}={\rm vol}_4 \ \lim_{r \rightarrow \infty}\Bigg[\, \frac{ r^4 f^{1/2}
\left(8 C^2+r f B'+2 r C C'+8 B f+r B f'\right)}{ L^5 h \sqrt{C^2+B f}} +\hspace{0.5in} %\hspace{0.5in}\nonumber \\
\nonumber \\
-\frac{r^4}{L^5}\sqrt{C^2+B f} (6- \psi^2)- \frac{2 r^4 f(r)}{L^5 h(r) }\sqrt{\frac{C(r)^2}{f(r)}+B(r)}\,\,\Bigg]\,.\label{Sren}
\eea
It is interesting to note that since we are always working with solutions with $\psi_1=0$,
the scalar piece can in fact be omitted if one desires.

%%%%%%%%%%%%%%%%%%%%%%%%%%%%%%%%%%%%%%%%%%%%%%%%%%%%%%%%%%%%%%%%%%%%%%%%%%%%%%%%%%%%%%%%%%%%%%%%%%%%%%%%%%%

\section{Superfluid Fraction}
\label{Sfraction}

In this section we present some details of the definition and computation of the superfluid fraction $\zeta$ for our solutions. We start with the renormalized action from the previous appendix and compute the boundary stress tensor and the boundary current by varying with respect to the {\em boundary} metric and the {\em boundary} components of the vector potential.
\beq
{\cal T}_{\mu\nu}=\frac{1}{\sqrt{-\gamma}}\frac{\delta S}{\delta \gamma^{\mu\nu}}
\;,\qquad
{\cal J}_{\mu}=\frac{1}{\sqrt{-\gamma}}\frac{\delta S}{\delta A^\mu}\,,
\eeq
where now $S=S_{IIB}+S_{\rm ct}$ with $S_{IIB}$ defined by (\ref{IIBac}) and $S_{\rm ct}$ defined by (\ref{BalCT}). In particular, the relations above are not tied to our ansatz. To compute the boundary stress tensor and current, we need to introduce coordinates on the boundary, and we will use Greek indices for them. After doing the variations, using our ansatz and going on shell on the bulk, the resulting stress tensor and current vanish in the strict $r \rightarrow \infty$ limit. This is consistent with the fact that they should be finite since we are using the renormalized action to compute them. The more interesting quantity is the boundary fluid stress tensor and the fluid current, which are defined in $AdS_5$ via
\beq
T_{\mu\nu}=\lim_{r \rightarrow \infty} r^2 {\cal T}_{\mu\nu}\;, \qquad J_\mu=\lim_{r \rightarrow \infty} r^2 {\cal J}_\mu\,.
\eeq
We are using units where $16 \pi G=1=L$ in this section.
Suppressing the details and restricting to our ansatz these quantities can be explicitly computed in terms of the boundary fall-offs of eqs. (\ref{falloff1})-(\ref{falloff4}) to be
\beq
T_{\mu\nu}=%2({\cal K}_{\mu\nu}-{\cal K}\gamma_{\mu\nu}+3 \gamma_{ab})-|\psi|^2\gamma_{\mu\nu}=
\left(
\begin{array}{cccc}
3f_4-B_4&4C_4&0&0 \\
4C_4&f_4-3B_4&0&0 \\
0&0&B_4+f_4&0 \\
0&0&0&B_4+f_4
\end{array}
\right)\;, \qquad
J_{\mu}={4\over3}\left(
\begin{array}{c}
A_{t,2} \\
A_{x,2}\\
0 \\
0
\end{array}
\right)\,.
\eeq
Now we follow the interpretation of \cite{Tisza} for these quantities in terms of a two-fluid model on the boundary, where one component is an ordinary (ideal) fluid and the other is a superfluid. First we can write these quantities suggestively in terms of $u_\mu=(-1,0,0,0)$ and $n_\mu=(0,1,0,0)$ as
\bea
T_{\mu\nu}=(\epsilon+P)\,u_\mu\, u_\nu +P\,\eta_{\mu\nu}-4B_4\,n_\mu\,n_\nu-8C_4\, u_{(\mu}\,n_{\nu)}\;,\qquad
J_\mu=\rho\, u_\mu-J_s\, n_\mu\,. \label{ourstress}
\eea
where
\bea
P\equiv f_4+B_4\;, \qquad \epsilon\equiv 3f_4-B_4\;, \qquad \rho\equiv-{4\over3}\,A_{t,2}\;, \qquad J_s\equiv-{4\over3}
\,A_{x,2}.
\eea
Note that what we have done is merely to rewrite the expressions covariantly in terms of the vectors $u_\mu$ and $n_\mu$. Another way to state the same thing is that (for example) the most general symmetric second rank tensor constructed from $u_\mu$ and $n_\mu$ will have to be a linear combination of $\eta_{\mu\nu}$, $u_\mu\,u_\nu\,$, $u_{(\mu}\,n_{\nu)}$ and  $n_\mu\, n_\nu$.

The two fluid model can be defined by the stress tensor
\bea
T_{\mu\nu}=(\epsilon_0+P_0)\,u_\mu\, u_\nu+P_0\, \eta_{\mu\nu}+\mu\,\rho_s\,v_\mu\, v_\nu\;, \qquad
J_\mu=\rho_n\, u_\mu+\rho_s\, v_\mu\,, \label{2fstress}
\eea
where the subscripts $n$ and $s$ stand for the normal and superfluid components of the charge density, with the total charge density $\rho=\rho_s+\rho_n$. Aside from the various thermodynamical state variables (whose precise interpretations will not be important to us, see \cite{Tisza}), we have also introduced the superfluid velocity $v_\mu$ that satisfies the constraint (``Josephson equation")
\bea
u^\mu \, v_\mu =-1.
\eea
%The subscripts $n$ and $s$ stand for the normal and superfluid components of the charge density, with the total charge density $\rho=\rho_s+\rho_n$.
The superfluid fraction is defined as
\bea
\zeta=\frac{\rho_s}{\rho}.
\eea
Our stress tensor (\ref{ourstress}) can be brought to the two-fluid form by defining $v_\mu$ as
\bea
v_\mu=u_\mu+\frac{B_4}{C_4}\,n_\mu\,.
\eea
This automatically satisfies $v_\mu u^\mu=-1$ as a consequence of $u_\mu u^\mu=-1$, and $n_\mu u^\mu=0$. %and $n_\mu n^\mu=1$. 
Rewriting our stress and current tensors (\ref{ourstress}) in these new variables we get the two-fluid form (\ref{2fstress}):
\bea
&&T_{\mu\nu}=(\epsilon+P+4 \ C_4^2/ B_4)\,u_\mu\, u_\nu +P\,\eta_{\mu\nu}-(4\,C_4^2/ B_4)\, v_\mu\, v_\nu\;, \\
&&J_\mu = (\rho + J_s\,C_4/B_4)\,u_\mu -(J_s\, C_4/B_4)\, v_\mu\,.
\eea
Reading off the superfluid fraction from this, we find that
\beq
\zeta=\frac{-(J_s\, C_4/B_4)}{\rho}=-\frac{A_{x,2}\,C_4}{A_{t,2}\,B_4}\,,
\eeq
where we have written the final result in terms of the fall-offs obtained directly from the solutions. This is the form we use for making the plots in figure 7.

%%%%%%%%%%%%%%%%%%%%%%%%%%%%%%%%%%%%%%%%%%%%%%%%%%%%%%%%%%%%%%%%%%%%%%%%%%%%%%%%%%%%%%%%%%%%%%%%%%%%%%%%%%%

\section{The Hairless Solution: Reissner-Nordstrom}
\label{RN}

In understanding the phase structure, it is important to keep in mind that we are interested in comparing the
free energy of the hairy black hole solution to that of Reissner-Nordstrom.
In the five dimensional IIB case, the Reissner-Nordstrom metric \cite{Gubser} can be given in terms of our ansatz (\ref{ModTisza4}) by
\bea
&&f(r)=1-\frac{1}{r^4}\Big(1+\frac{4\mu^2}{9}\Big)+\frac{4\mu^2}{9r^6}\,,\quad
A_t=\mu\Big(1-\frac{1}{r^2}\Big)\,, \\
&&h=1\,,\quad
B=1\,,\quad
C=0\,,\quad
A_x=0\,,\quad
\psi=0\,.
\eea
In this notation, the curvature invariants studied in section \ref{zeroT} take the form
\bea
&&R=-20f-r\left(10f'+rf''\right)\,, \\
&&R_{abcd}\,R^{abcd}=40f^2+4r\,f\left(10f'+r\,f''\right)+r^2\left[22(f')^2+8r\,f'\,f''+r^2(f'')^2\right]\,.
\label{riemform}
\eea
All these expressions are obtained after all the necessary rescalings: we have set $16\pi G=L=r_H=1$. The Hawking temperature now takes the form
\beq
T_H= \frac{1-2\mu^2/9}{\pi}\,,
\eeq
as can be determined by the periodicity of the Euclidean section. The renormalized on-shell action that we determined
before takes a simple form for this solution:
\beq
S_{ren}=1+4\mu^2/9\,.
\eeq
We will compare the free energies of the hairy and hairless cases at the same $T/\mu$ to determine which one is the favored phase.

% ==========================================================================
%
%%%%%%%%%%%%%%%%%%%%%%%%%%%%%%%%%%%%%%%%%%%%%%%%%%%%%%%%%%%%%%%%%%%%%%%%%%%%
%                      REFERENCES                            %
%%%%%%%%%%%%%%%%%%%%%%%%%%%%%%%%%%%%%%%%%%%%%%%%%%%%%%%%%%%%%%%%%%%%%%%%%%%%
%\newpage
%\bibliography{metasusy}

\begin{thebibliography}{19}        %here 19 is the widest mark...
%-----Type it \bibitem[how it is marked]{how we call it}Authors,
%-----Citations are then made by \cite{how we call it} in text
%-----\bibitem without [how it is denoted] is numbered 1,2,3....


%
%\cite{Gubser:2008px}
\bibitem{Gubser:2008px}
  S.~S.~Gubser,
  ``Breaking an Abelian gauge symmetry near a black hole horizon,''
  Phys.\ Rev.\  D {\bf 78} (2008) 065034
  [arXiv:0801.2977 [hep-th]].
  %%CITATION = PHRVA,D78,065034;%%
%
\bibitem{Maldacena}
  J.~M.~Maldacena,
  ``The large N limit of superconformal field theories and supergravity,''
  Adv.\ Theor.\ Math.\ Phys.\  {\bf 2}, 231 (1998)
  [Int.\ J.\ Theor.\ Phys.\  {\bf 38}, 1113 (1999)]
  [arXiv:hep-th/9711200].
  %%CITATION = IJTPB,38,1113;%%
%
%\cite{Gubser:1998bc}
\bibitem{GKP}
  S.~S.~Gubser, I.~R.~Klebanov and A.~M.~Polyakov,
  ``Gauge theory correlators from non-critical string theory,''
  Phys.\ Lett.\  B {\bf 428}, 105 (1998)
  [arXiv:hep-th/9802109].
  %%CITATION = PHLTA,B428,105;%%
%
\bibitem{Witten}
  E.~Witten,
  ``Anti-de Sitter space and holography,''
  Adv.\ Theor.\ Math.\ Phys.\  {\bf 2}, 253 (1998)
  [arXiv:hep-th/9802150].
  %%CITATION = 00203,2,253;%%
%
%\cite{weinberg}
\bibitem{weinberg}
S.~Weinberg, {\it The quantum theory of fields, Vol II}, CUP, 1996.
%
%\cite{Hartnoll:2008vx}
\bibitem{HHH1}
  S.~A.~Hartnoll, C.~P.~Herzog and G.~T.~Horowitz,
  ``Building a Holographic Superconductor,''
  Phys.\ Rev.\ Lett.\  {\bf 101}, 031601 (2008)
  [arXiv:0803.3295 [hep-th]].
  %%CITATION = PRLTA,101,031601;%%
%
%\cite{Hartnoll:2009sz}
\bibitem{Hartnoll:2009sz}
  S.~A.~Hartnoll,
  ``Lectures on holographic methods for condensed matter physics,''
  Class.\ Quant.\ Grav.\  {\bf 26} (2009) 224002
  [arXiv:0903.3246 [hep-th]].
  %%CITATION = CQGRD,26,224002;%%
%
%\cite{Herzog:2009xv}
\bibitem{Herzog:2009xv}
  C.~P.~Herzog,
  ``Lectures on Holographic Superfluidity and Superconductivity,''
  J.\ Phys.\ A  {\bf 42} (2009) 343001
  [arXiv:0904.1975 [hep-th]].
  %%CITATION = JPAGB,A42,343001;%%
%
%\cite{Horowitz:2010gk}
\bibitem{Horowitz:2010gk}
  G.~T.~Horowitz,
  ``Introduction to Holographic Superconductors'',
  arXiv:1002.1722 [hep-th].
  %%CITATION = ARXIV:1002.1722;%%
%
\bibitem{GubRoch2}
  S.~S.~Gubser and F.~D.~Rocha,
  ``The gravity dual to a quantum critical point with spontaneous symmetry
  breaking,''
  Phys.\ Rev.\ Lett.\  {\bf 102}, 061601 (2009)
  [arXiv:0807.1737 [hep-th]].
  %%CITATION = PRLTA,102,061601;%%

\bibitem{Gubser}
  S.~S.~Gubser, C.~P.~Herzog, S.~S.~Pufu and T.~Tesileanu,
  ``Superconductors from Superstrings,''
  Phys.\ Rev.\ Lett.\  {\bf 103}, 141601 (2009)
  [arXiv:0907.3510 [hep-th]].
  %%CITATION = PRLTA,103,141601;%%

%\cite{Gauntlett:2009dn}
\bibitem{Gaunt}
  J.~P.~Gauntlett, J.~Sonner and T.~Wiseman,
  ``Holographic superconductivity in M-Theory,''
  Phys.\ Rev.\ Lett.\  {\bf 103} (2009) 151601
  [arXiv:0907.3796 [hep-th]].
  %%CITATION = PRLTA,103,151601;%%

%\cite{Ammon:2009fe}
\bibitem{Ammon:2009fe}
  M.~Ammon, J.~Erdmenger, M.~Kaminski and P.~Kerner,
  ``Superconductivity from gauge/gravity duality with flavor,''
  Phys.\ Lett.\  B {\bf 680} (2009) 516
  [arXiv:0810.2316 [hep-th]];
  %%CITATION = PHLTA,B680,516;%%
  ``Flavor Superconductivity from Gauge/Gravity Duality,''
  JHEP {\bf 0910} (2009) 067
  [arXiv:0903.1864 [hep-th]].
  %%CITATION = JHEPA,0910,067;%%

\bibitem{HHH2}
  S.~A.~Hartnoll, C.~P.~Herzog and G.~T.~Horowitz,
  ``Holographic Superconductors,''
  JHEP {\bf 0812}, 015 (2008)
  [arXiv:0810.1563 [hep-th]].
  %%CITATION = JHEPA,0812,015;%%
%
\bibitem{Basu}
  P.~Basu, A.~Mukherjee and H.~H.~Shieh,
  ``Supercurrent: Vector Hair for an AdS Black Hole,''
  Phys.\ Rev.\  D {\bf 79}, 045010 (2009)
  [arXiv:0809.4494 [hep-th]].
  %%CITATION = PHRVA,D79,045010;%%
%\cite{Herzog:2008he}
%
\bibitem{Herzog}
  C.~P.~Herzog, P.~K.~Kovtun and D.~T.~Son,
  ``Holographic model of superfluidity,''
  Phys.\ Rev.\  D {\bf 79}, 066002 (2009)
  [arXiv:0809.4870 [hep-th]].
  %%CITATION = PHRVA,D79,066002;%%
%\cite{Witten:1998qj}
%
%\cite{Faulkner:2010gj}
\bibitem{Faulkner:2010gj}
  T.~Faulkner, G.~T.~Horowitz, M.~M.~Roberts,
  ``Holographic quantum criticality from multi-trace deformations,''
  [arXiv:1008.1581 [hep-th]].
%
%\cite{Keranen:2009re}
\bibitem{Keranen:2009re}
  V.~Keranen, E.~Keski-Vakkuri, S.~Nowling and K.~P.~Yogendran,
  ``Inhomogeneous Structures in Holographic Superfluids: II. Vortices,''
  Phys.\ Rev.\  D {\bf 81} (2010) 126012
  [arXiv:0912.4280 [hep-th]].
  %%CITATION = PHRVA,D81,126012;%%
%
%\cite{Arean:2010xd}
\bibitem{Daniel}
  D.~Arean, M.~Bertolini, J.~Evslin and T.~Prochazka,
  ``On Holographic Superconductors with DC Current,''
  JHEP {\bf 1007} (2010) 060
  [arXiv:1003.5661 [hep-th]].
  %%CITATION = JHEPA,1007,060;%%
%
%\cite{Tinkham}
\bibitem{Tinkham}
M.~Tinkham,
{\it Introduction to Superconductivity},
2nd edition, Dover: New York (1996).
%


\bibitem{Tisza}
  J.~Sonner and B.~Withers,
  ``A gravity derivation of the Tisza-Landau Model in AdS/CFT,''
  arXiv:1004.2707 [hep-th].
  %%CITATION = ARXIV:1004.2707;%%
%

%\cite{Arean:2010zw}
\bibitem{Daniel2}
  D.~Arean, P.~Basu and C.~Krishnan,
  ``The Many Phases of Holographic Superfluids,''
  JHEP {\bf 1010}, 006 (2010)
  [arXiv:1006.5165 [hep-th]].
  %%CITATION = JHEPA,1010,026;%%



%
%\cite{Gubser:2009gp}
\bibitem{GubRoch}
  S.~S.~Gubser, S.~S.~Pufu and F.~D.~Rocha,
  ``Quantum critical superconductors in string theory and M-theory,''
  Phys.\ Lett.\  B {\bf 683} (2010) 201
  [arXiv:0908.0011 [hep-th]].
  %%CITATION = PHLTA,B683,201;%%
%

\bibitem{HoroRob}
  G.~T.~Horowitz and M.~M.~Roberts,
  ``Zero Temperature Limit of Holographic Superconductors,''
  JHEP {\bf 0911}, 015 (2009)
  [arXiv:0908.3677 [hep-th]].
  %%CITATION = JHEPA,0911,015;%%

%
%\cite{Girardello:1999bd}
\bibitem{Girardello:1999bd}
  L.~Girardello, M.~Petrini, M.~Porrati and A.~Zaffaroni,
  ``The supergravity dual of N = 1 super Yang-Mills theory,''
  Nucl.\ Phys.\  B {\bf 569} (2000) 451
  [arXiv:hep-th/9909047].
  %%CITATION = NUPHA,B569,451;%%
%
%\cite{Distler:1999tr}
\bibitem{Distler:1999tr}
  J.~Distler and F.~Zamora,
  ``Chiral symmetry breaking in the AdS/CFT correspondence,''
  JHEP {\bf 0005} (2000) 005
  [arXiv:hep-th/9911040].
  %%CITATION = JHEPA,0005,005;%%

%\cite{Bobev:2010ib}
\bibitem{Bobev:2010ib}
  N.~Bobev, N.~Halmagyi, K.~Pilch and N.~P.~Warner,
  ``Supergravity Instabilities of Non-Supersymmetric Quantum Critical Points,''
  arXiv:1006.2546 [hep-th].
  %%CITATION = ARXIV:1006.2546;%%

%\cite{Gauntlett:2009bh}
\bibitem{Gauntlett:2009bh}
  J.~P.~Gauntlett, J.~Sonner and T.~Wiseman,
  ``Quantum Criticality and Holographic Superconductors in M-theory,''
  JHEP {\bf 1002} (2010) 060
  [arXiv:0912.0512 [hep-th]].
  %%CITATION = JHEPA,1002,060;%%

%
%\cite{Balasubramanian:1999re}
\bibitem{Bala}
  V.~Balasubramanian and P.~Kraus,
  ``A stress tensor for anti-de Sitter gravity,''
  Commun.\ Math.\ Phys.\  {\bf 208}, 413 (1999)
  [arXiv:hep-th/9902121].
  %%CITATION = CMPHA,208,413;%%


\end{thebibliography}

\end{document}